\documentstyle[preprint,prc,aps,eqsecnum,epsfig]{revtex}
\tighten
\begin{document}
\draft

\title{Charge symmetry breaking of the nucleon-nucleon interaction:
$\rho$-$\omega$ mixing {\it versus} nucleon mass splitting}

\author{R. Machleidt$^{1,\!\!}$~\footnote{e-mail address: machleid@uidaho.edu}
     and H. M\"uther$^{2,\!\!}$~\footnote{e-mail address: 
 herbert.muether@uni-tuebingen.de}}

\address{$^1$Department of Physics, University of Idaho, Moscow,
Idaho 83844, U. S. A.\\
$^2$Institut f\"ur Theoretische Physik, Universit\"at T\"ubingen,
D-72076 T\"ubingen, Germany}

\date{\today}

\maketitle

%\tableofcontents

\begin{abstract}
We investigate three models for the charge symmetry breaking (CSB)
of the nucleon-nucleon ($NN$) interaction 
(based upon $\rho$-$\omega$ mixing, 
nucleon mass splitting, and phenomenology)
that all reproduce the empirical value for the CSB of
the $^1S_0$ scattering length ($\Delta a_{CSB}$) accurately. 
We reveal that these models make very different
predictions for CSB in $^3P_J$ waves and examine
the impact of this on some
observable quantities of $A\geq 3$ nuclear systems.
It turns out that the $^3$H-$^3$He binding energy difference
is essentially ruled by $\Delta a_{CSB}$ and 
not very sensitive to CSB from $P$ waves.
However, 
the Coulomb displacement energies (which are the subject
of the Nolen-Schiffer anomaly) receive about 50\% of their
CSB contribution from $NN$ partial waves beyond $^1S_0$.
Consequently, the predictions by the various CSB models
differ here substantially (10-20\%).
Unfortunately, the evaluation of the leading Coulomb contributions
carry a large uncertainty
such that no discrimination between the competing CSB models 
can presently be made.
To decide the issue we suggest to look into nuclear few-body 
reactions that are sensitive to CSB of the nuclear force.
\end{abstract}


%\twocolumn 

\section{Introduction}

By definition, {\it charge independence} is invariance under any 
rotation in isospin space. 
A violation of this symmetry is referred to as charge dependence
or charge independence breaking (CIB).
{\it Charge symmetry} is invariance under a rotation by 180$^0$ about the
$y$-axis in isospin space if the positive $z$-direction is associated
with the positive charge.
The violation of this symmetry is known as charge symmetry breaking (CSB).
Obviously, CSB is a special case of charge dependence.

CIB of the strong $NN$ interaction means that,
in the isospin $T=1$ state, the
proton-proton ($T_z=+1$), 
neutron-proton ($T_z=0$),
or neutron-neutron ($T_z=-1$)
interactions are (slightly) different,
after electromagnetic effects have been removed.
CSB of the $NN$ interaction refers to a difference
between proton-proton ($pp$) and neutron-neutron ($nn$)
interactions, only. 

Charge asymmetry is seen most clearly in the $^1S_0$ scattering
length.
The latest empirical values of the neutron-neutron ($nn$) singlet 
effective range
parameters (corrected for electromagnetic effects) are
\begin{equation}
a^N_{nn}=-18.9\pm 0.4 \mbox{ fm} \; \cite{How98,Gon99}, \hspace*{0.2cm}
    r^N_{nn} = 2.75\pm 0.11 \mbox{ fm} \; \cite{MNS90}.
\end{equation}
This should be compared to the corresponding proton-proton ($pp$) 
values~\cite{MNS90}:
\begin{equation}
a^N_{pp}=-17.3\pm 0.4 \mbox{ fm}, \hspace*{.2cm} 
     r^N_{pp}=2.85\pm 0.04 \mbox{ fm}.
\end{equation}
The implication is that the singlet effective range parameters break
charge-symmetry by the following amounts,
\begin{eqnarray}
\Delta a_{CSB}& \equiv& a_{pp}^N-a_{nn}^N = 1.6\pm 0.6 \mbox{ fm}, 
\label{eq_CSBlep1}
\\
\Delta r_{CSB}& \equiv& r_{pp}^N-r_{nn}^N = 0.10\pm 0.12 \mbox{ fm}. 
\label{eq_CSBlep2}
\end{eqnarray}

The current understanding is 
that---on a fundamental level---the 
charge dependence of nuclear forces is due to a
difference between the up and down quark masses and electromagnetic 
interactions among the quarks. 
As a consequence of this---on the hadronic level---major
causes of CIB are mass differences between hadrons of the same
isospin multiplet, meson mixing, and irreducible meson-photon exchanges.
For recent reviews on charge dependence, 
see Refs.~\cite{MNS90,Mac89,MO95}.

\begin{figure}
\vspace*{-3cm}
\hspace*{-1.cm}
\epsfig{file=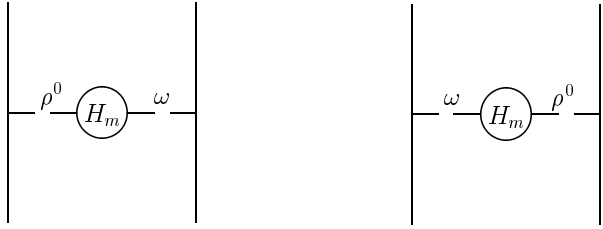,width=15cm}

\vspace*{-9cm}
\caption{$\rho^0$-$\omega$ exchange contributions to the nuclear force
that violate charge symmetry.}
\label{fig_rom}
\end{figure}

Neutral mesons with the same spin and parity, but different isospin,
may mix due to the up-down quark mass difference and electromagnetic interactions;
that is, the Hamiltonian responsible for the mixing, $H_m$, has
a strong and an electromagnetic part,
\begin{equation}
H_m = H_{str} + H_{em} \, .
\label{eq_hm1}
\end{equation}
The most prominent case is $\rho^0$-$\omega$ mixing that is observed in the
annihilation process $e^+ e^- \rightarrow \pi^+ \pi^-$ from which
the mixing matrix element has been extracted~\cite{CB87} to be
\begin{equation}
\langle \rho^0 | H_m | \omega \rangle = - 0.00452 \pm 0.0006 
\mbox{ GeV$^2$.}
\label{eq_hm2}
\end{equation}
The charge-asymmetric nuclear force
created by this process is displayed in Fig.~\ref{fig_rom}.
Coon and Barrett showed~\cite{CB87} that this mechanism alone
can explain the entire empirical CSB of the singlet scattering length,
Eq.~(\ref{eq_CSBlep1}).
Other examples of meson mixing are 
$\pi$-$\eta$
 and $\pi$-$\eta'$
mixing which, however, generate only negligible effects~\cite{CS82}.

In recent years, 
the process of Fig.~\ref{fig_rom} has been subjected to some criticism.
Note that the matrix element, Eq.~(\ref{eq_hm2}),
is extracted on-shell, i.~e., for $k^2=m_\rho^2$, where $k$ denotes the
four-momentum of the $\rho$ meson and $m_\rho$ the $\rho$ mass.
However, in the $NN$ interaction, Fig.~\ref{fig_rom}, the relevant $k^2$
are space-like (less than zero).
Using a quark model for $H_{str}$,
Goldman, Henderson, and Thomas~\cite{GHT92} find a substantial $k^2$
dependence which is such
that the contributions of Fig.~\ref{fig_rom}
would nearly vanish. Similar results were reported in subsequent
papers~\cite{PW93,KTW93,Con94}.
On the other hand, Miller~\cite{MO95} and Coon and coworkers~\cite{CMR97}
have advanced counter-arguments that would restore the traditional role
of $\rho$-$\omega$ exchange.
The issue is unresolved.
Informative summaries of the controversial points of view can be found in
Refs.~\cite{MO95,Con97,CS00}.

We now turn to another basic source for CSB of the nuclear force,
namely, nucleon mass splitting.
The most trivial consequence of nucleon mass splitting is a difference
in the kinetic energies: for the heavier neutrons, the kinetic energy
is smaller than for protons. This raises the magnitude of the
$nn$ scattering length by 0.25 fm as compared to $pp$.
Besides this, nucleon mass splitting has an impact on all meson-exchange
diagrams that contribute to the nuclear force.
For example, there are
the one-boson exchange (OBE) diagrams, Fig.~\ref{fig_mn1},
which are affected by only a negligible amount. 
However, a sophisticated and realistic meson model for the 
nuclear force~\cite{MHE87}
goes beyond single meson exchange and includes
irreducible diagrams of
two-boson exchange which generate substantial CSB from nucleon mass splitting.
The major part of the CSB effect comes from diagrams of
$2\pi$ exchange with $N\Delta$ intermediate states, Fig.~\ref{fig_mn2}. 
In fact, these diagrams can fully explain
the empirical CSB splitting of the singlet 
scattering length~\cite{CN96,LM98a}.

Finally, for reasons of completeness, we mention that irreducible
diagrams of $\pi$ and $\gamma$ exchange between
two nucleons create a charge-dependent nuclear force.
Recently, these contributions have been calculated to
leading order in chiral perturbation theory~\cite{Kol98}.
It turns out that to this order the $\pi\gamma$ force is
charge-symmetric (but does break charge independence).

The bottom line then is that we have two CSB mechanisms in hand,
each powerful enough to fully explain the charge asymmetry seen in the
singlet scattering length.
This state of affairs challenges the question:
{\it Which of the two mechanisms is nature really using?}
To answer the question, the $^1S_0$ state at zero energy is obviously
of no use since this can be described equally well by both
mechanisms. Thus, the answer, if any, can only come from energies $T_{lab}>0$
and/or states with $L>0$. It is quite possible that 
the predictions by the two mechanisms are very different 
in states other than $^1S_0$. The $2\pi$ exchange contribution to the nuclear
force (where the CSB effect due to nucleon mass splitting mainly
comes from) is chiefly a central force of intermediate range,
while vector meson exchange (involved in $\rho$-$\omega$ mixing)
creates a large spin-orbit component and a short-ranged central force.
This may create large differences in triplet $P$ waves where
the signature of spin-orbit forces is most pronounced.
If large differences between the models occur for $L>0$,
then each mechanism may have a characteristic signature on
observables that are sensitive to CSB from $P$ and higher partial waves.
In Ref.~\cite{MPM99} it was found that about 50\%
of the CSB contribution to the Nolen-Schiffer (NS)
anomaly~\cite{NS69} comes from the $L>0$ two-nucleon states.
Thus, there is the possibility that the differences between competing CSB models
may show up as differences in the predictions for the NS anomaly.
Other observables may be considered, too.

It is the purpose of this paper to look into this issue.
In particular, we want to find out if it is possible to discriminate
between the two models for CSB based upon their predictions
for observable quantities, other than the singlet scattering length.

Besides the two CSB mechanisms discussed, we will include in our study also
a phenomenological model for CSB, namely, the Argonne $V_{18}$
potential~\cite{WSS95}.
The three models will be introduced in Sec.~II and applied
to the $^3$H-$^3$He binding energy difference and the NS anomaly
in Sec.~III.
Sec.~IV concludes the paper.

\section{Models for charge symmetry breaking of the two-nucleon interaction}

\subsection{$\rho$-$\omega$ mixing}
Here we will evaluate the charge asymmetric nuclear force generated
by $\rho^0$-$\omega$ mixing shown in Fig.~\ref{fig_rom} \cite{MSC75}.

The coupling of $\rho$ and $\omega$ mesons 
to nucleons is described by the following Lagrangians: 
\begin{eqnarray}
{\cal L}_{\omega NN}&=& -g_{\omega}\bar{\psi}\gamma^{\mu}\psi
                       \varphi^{(\omega)}_{\mu}
\; ,
\label{eq_omegaNN}
\\
{\cal L}_{\rho NN} &=& -g_{\rho}
                           \bar{\psi}  \gamma^\mu  
                           \mbox{\boldmath $\tau$} \psi
                           \cdot 
                           \mbox{\boldmath $\varphi$}^{(\rho)}_\mu
                  - \frac{f_\rho}{4M_p}
                           \bar{\psi}  \sigma^{\mu\nu}  
                           \mbox{\boldmath $\tau$} \psi
                           \cdot 
                           (\partial_\mu
                           \mbox{\boldmath $\varphi$}^{(\rho)}_\nu
                           -\partial_\nu
                           \mbox{\boldmath $\varphi$}^{(\rho)}_\mu)
\; ,
\label{eq_rhoNN}
\end{eqnarray}
where $\psi$ denotes nucleon fields, $\varphi$ meson fields, and
$\tau_{i}$ ($i=1,2,3$) are the usual Pauli matrices
describing isospin $\frac12$; specifically, 
$\tau_3 | p \rangle = + | p \rangle$
and $\tau_3 | n \rangle = - | n \rangle$
with $ | p \rangle$ a proton state
and $ | n \rangle$  a neutron state.
$M_p$ is the proton mass which is used as scaling mass in the $\rho NN$
Lagrangian to make $f_\rho$ dimensionless. To avoid the creation
of unmotivated charge dependence, the scaling mass $M_p$ is
used in $pp$ as well as $nn$ scattering.

The first Feynman diagram of Fig.~\ref{fig_rom}
leads to the following $\rho$-$\omega$ potential,
\begin{eqnarray}
\langle {\bf  q'} \lambda_{1}'\lambda_{2}'
|{V}_{\rho\omega}^{(1)}|
{\bf  q}\lambda_{1}\lambda_{2}\rangle
 &  = &
 - \frac{1}{(2\pi)^3} \;
\sqrt{\frac{M}{E'}} \;
\sqrt{\frac{M}{E}} \;
\langle \rho^0|H_m|\omega\rangle \;
  g_\rho \; g_\omega 
\nonumber \\ & & \times
\bar{u}({\bf  q'},\lambda_{1}')  
\; \tau_3
\left[
\gamma_{\mu} 
+\frac{\kappa_{\rho}}{2M_p}
\sigma_{\mu\nu}i(q'-q)^{\nu}
\right]
 u({\bf  q},\lambda_{1})
\nonumber \\ & & \times
\bar{u}({\bf  -q'},\lambda_{2}')
  \; \gamma^{\mu} \; 
u({\bf  -q},\lambda_{2})
\nonumber \\  &  &  \times 
\frac{
{\cal F}_\rho[({\bf  q}'-{\bf  q})^2] \;
{\cal F}_\omega[({\bf  q}'-{\bf  q})^2]}
{[({\bf  q'-q})^{2}+m_{\rho}^{2}] 
           [({\bf  q'-q})^{2}+m_{\omega}^{2}]} 
\: ,
\label{eq_rom1}
\end{eqnarray}
where 
$M$ denotes the relevant nucleon mass (i.~e., 
$M=M_p$ in $pp$ scattering
and $M=M_n$ in $nn$ scattering),
$m_\alpha$ are meson masses, and
$\kappa_\rho\equiv f_\rho/g_\rho$.
We are working here in the two-nucleon c.m.\ frame
and use the helicity formalism.
The helicity $\lambda_i$ of nucleon $i$ is defined
as the eigenvalue of the helicity operator 
$\frac12 \bbox{\sigma}_i \cdot {\bf p}_i/|{\bf p}_i|$ which is 
$\pm \frac12$. 
In-coming nucleon 1 carries helicity $\lambda_1$ 
and momentum ${\bf q}$ 
and in-coming nucleon 2 carries helicity $\lambda_2$ 
and momentum ${\bf -q}$;
the out-going nucleons have
$\lambda_1'$, ${\bf q'}$ and $\lambda_2'$, ${\bf -q'}$.
The above `quasi-potential' is defined in the framework
of the relativistic, three-dimensional 
Blankenbecler-Sugar (BbS) equation~\cite{BS66}
which is a reduced version of the four-dimensional 
relativistic Bethe-Salpeter equation~\cite{SB51}.
In the BbS formalism, the four-momentum transfer between
the two nucleons is $(q'-q)^\mu=(0,{\bf  q'-q})$.
The square-root factor
$M/\sqrt{E'E}$
(with $E\equiv \sqrt{M^2+{\bf q}^2}$
and $E'\equiv \sqrt{M^2+{\bf q'}^2}$)
makes it possible to cast the BbS equation
into a form that is identical to the conventional Lippmann-Schwinger
equation.
The Dirac spinors in helicity representation are given by
\begin{eqnarray}
u({\bf  q},\lambda_1)&=&\sqrt{\frac{E+M}{2M}}
\left( \begin{array}{c}
       1\\
       \frac{2\lambda_1 |{\bf q}|}{E+M}
       \end{array} \right)
|\lambda_1\rangle
\: ,
\\
u(-{\bf  q},\lambda_2)&=&\sqrt{\frac{E+M}{2M}}
\left( \begin{array}{c}
       1\\
       \frac{2\lambda_2 |{\bf q}|}{E+M}
       \end{array} \right)
|\lambda_2\rangle
\: ,
\end{eqnarray}
with normalization
\begin{equation}
\bar{u}({\bf  q},\lambda) u({\bf  q},\lambda)=1,
\end{equation}
where $\bar{u}\equiv u^{\dagger}\gamma^{0}$.
The above amplitude, includes form factors
\begin{equation}
{\cal F}_\alpha[({\bf  q}'-{\bf  q})^2]=
\frac{\Lambda^2_\alpha-m^2_\alpha}
{\Lambda^2_\alpha+({\bf  q}'-{\bf  q})^2} 
\label{eq_ff}
\end{equation}
with $m_\alpha$ the mass of the meson involved and  
$\Lambda_\alpha$ the so-called cutoff mass.
For more details concerning the formalism and the explicit evaluation
of quasi-potentials of this kind, see Appendix E of Ref.~\cite{MHE87}.

The second Feynman diagram of Fig.~\ref{fig_rom} yields the same as the first 
one and, so, the entire $\rho$-$\omega$ potentials is
\begin{equation}
V_{\rho\omega} 
= 2 
V_{\rho\omega}^{(1)}
\end{equation}
with
$V_{\rho\omega}^{(1)}$ given by Eq.~(\ref{eq_rom1}).
Since 
$\langle \rho^0|H_m|\omega\rangle$
is negative,
$V_{\rho\omega}$ 
is repulsive for $pp$ scattering 
and attractive for $nn$ scattering with the magnitude of
$V_{\rho\omega}$ 
essentially the same in both cases.

When constructing CSB $NN$ potentials, one starts with
the $pp$ potential since there are many $pp$ data 
(and $pp$ phase shift analyses)
of high quality to constrain the $pp$ potential.
In this work, we use the CD-Bonn $pp$ potential~\cite{Mac00}
which reproduces the world $pp$ data below 350 MeV
lab.\ energy available in the year of 2000 with the perfect
$\chi^2$/datum of 1.01.
The $nn$ potential is then fabricated by adding
to the $pp$ potential a difference-potential that contains
the entire difference between $nn$ and $pp$ due to $\rho$-$\omega$
exchange, which is
\begin{equation}
\Delta V_{\rho\omega} 
=
V_{\rho\omega}(nn) 
-
V_{\rho\omega}(pp) 
\approx
-4\check{V}_{\rho\omega}^{(1)}(M_n)
\label{eq_delv}
\end{equation}
where
$\check{V}_{\rho\omega}^{(1)}$ is 
$V_{\rho\omega}^{(1)}$ [Eq.~(\ref{eq_rom1})]
with the $\tau_3$ operator replaced by 1. 
Note that---strictly speaking--- $V_{\rho\omega}(nn)$ is to be
evaluated with $M=M_n$ and
$V_{\rho\omega}(pp)$ with $M=M_p$.
However, if we wish to subsume both terms into one, then
we have to use the same mass for both for which we choose $M=M_n$.
We have tested this approximation and found that it affects the singlet
scattering length by $10^{-4}$ fm.

To obtain a convenient expression for
$\Delta V_{\rho\omega}$, 
we make use of 
the identity
\begin{equation}
\frac{1}
{[({\bf  q'-q})^{2}+m_{\rho}^{2}] 
           [({\bf  q'-q})^{2}+m_{\omega}^{2}]} 
=
\frac{1}
{m_\omega^2 - m_\rho^2}
\left[
\frac{1}{({\bf  q'-q})^{2}+m_{\rho}^{2}} 
-
\frac{1}{({\bf  q'-q})^{2}+m_{\omega}^{2}} 
\right] \, ,
\end{equation}
which allows us to write 
$\Delta V_{\rho\omega}$ 
in terms of the difference of two expressions
each of which resembles single meson exchange; namely,
\begin{equation}
\Delta V_{\rho\omega} 
=
\Delta V_{\rho\omega}^{(\rho)}
-
\Delta V_{\rho\omega}^{(\omega)}
\end{equation}
with
\begin{eqnarray}
\langle {\bf  q'} \lambda_{1}'\lambda_{2}'
|\Delta V_{\rho\omega}^{(\rho)}|
{\bf  q}\lambda_{1}\lambda_{2}\rangle
 &  = &
  \frac{4}{(2\pi)^3} \;
\sqrt{\frac{M}{E'}} \;
\sqrt{\frac{M}{E}} \;
\frac{\langle \rho^0|H_m|\omega\rangle}
{m_\omega^2 - m_\rho^2} \;
  g_\rho \; g_\omega 
\nonumber \\ & & \times
\bar{u}({\bf  q'},\lambda_{1}')  
\left[
\gamma_{\mu} 
+\frac{\kappa_{\rho}}{2M_p}
\sigma_{\mu\nu}i(q'-q)^{\nu}
\right]
 u({\bf  q},\lambda_{1})
\nonumber \\ & & \times
\bar{u}({\bf  -q'},\lambda_{2}')
  \; \gamma^{\mu} \;
u({\bf  -q},\lambda_{2})
\nonumber \\ & & \times
\frac{
{\cal F}_\rho[({\bf  q}'-{\bf  q})^2] \;
{\cal F}_\omega[({\bf  q}'-{\bf  q})^2]}
{({\bf  q'-q})^{2}+m_{\rho}^{2}} 
\: ,
\label{eq_rom3}
\end{eqnarray}
and
\begin{eqnarray}
\langle {\bf  q'} \lambda_{1}'\lambda_{2}'
|\Delta V_{\rho\omega}^{(\omega)}|
{\bf  q}\lambda_{1}\lambda_{2}\rangle
 &  = &
  \frac{4}{(2\pi)^3} \;
\sqrt{\frac{M}{E'}} \;
\sqrt{\frac{M}{E}} \;
\frac{\langle \rho^0|H_m|\omega\rangle}
{m_\omega^2 - m_\rho^2} \;
  g_\rho \; g_\omega 
\nonumber \\ & & \times
\bar{u}({\bf  q'},\lambda_{1}')  
\left[
\gamma_{\mu} 
+\frac{\kappa_{\rho}}{2M_p}
\sigma_{\mu\nu}i(q'-q)^{\nu}
\right]
 u({\bf  q},\lambda_{1})
\nonumber \\ & & \times
\bar{u}({\bf  -q'},\lambda_{2}')
  \; \gamma^{\mu} \;
u({\bf  -q},\lambda_{2})
\nonumber \\ & & \times
\frac{
{\cal F}_\rho[({\bf  q}'-{\bf  q})^2] \;
{\cal F}_\omega[({\bf  q}'-{\bf  q})^2]}
{({\bf  q'-q})^{2}+m_{\omega}^{2}} 
\: .
\label{eq_rom4}
\end{eqnarray}

For the masses involved, we use~\cite{PDG98},
\begin{eqnarray}
M_p & = &  938.27231    
   \mbox{ MeV,} \\
M   & = &  M_n  = 939.56563  
   \mbox{ MeV,} \\
m_\rho & = & 769.9 
   \mbox{ MeV,} \\
m_\omega & = &  781.94     
   \mbox{ MeV,} 
\end{eqnarray}
and $\Lambda_\rho = \Lambda_\omega = 1.4$ GeV.

We choose
for the meson-nucleon coupling constants, 
\begin{eqnarray}
\frac{g^2_\rho}{4\pi} & = & 0.84 \, , \\ 
\kappa_\rho & = & 6.1  \, , \\
\frac{g^2_\omega}{4\pi} & = & 10 \, , 
\label{eq_gomega}
\end{eqnarray}
and for the mixing matrix element,
\begin{equation}
\langle \rho^0 | H_m | \omega \rangle = - 0.00402 \mbox{ GeV$^2$,}
\label{eq_hm4}
\end{equation}
to obtain
\begin{equation}
\Delta a_{CSB} = 1.508 \mbox{ fm.} 
\end{equation}
The above mixing matrix element is consistent with the empirical value,
Eq.~(\ref{eq_hm2}), and the $\rho$ parameters are identical to the ones
used in the Bonn Full Model~\cite{MHE87}.
Concerning the $\omega$, the Bonn model uses
$g^2_\omega/4\pi = 20$ which, however, would generate too much CSB
when applied in the above $\rho$-$\omega$ potential;
therefore our choice Eq.~(\ref{eq_gomega}).
This choice could be justified with the argument 
that part of the $\omega$ contribution in meson-theoretic potentials
may be just a parametrization of short-ranged repulsion that is actually
due to quark-gluon exchange.

To check our calculations,
we have made a comparison with the results by
Coon and Barrett~\cite{CB87}.
Note that these authors use very different vector-meson
coupling constants as compared to ours. 
In terms of our convention for the coupling constants, 
Eqs.~(\ref{eq_omegaNN}) and
(\ref{eq_rhoNN})~\cite{foot1}, 
Coon and Barrett use
$g^2_\rho/4\pi  =  0.6$,  
$\kappa_\rho  =  3.7$, 
$g^2_\omega/4\pi  =  5.25$, 
and
$\kappa_\omega  = -0.12$. 
There are also other differences, like, we use the the full
relativistic Feynman amplitudes for $\rho^0$-$\omega$ exchange,
Eq.~(\ref{eq_rom1}), while in Ref.~\cite{CB87} the nonrelativistic
approximation is applied.
Moreover, Coon and Barrett use the Reid potential~\cite{Rei68} as their basic $pp$
potential whereas we use the $pp$ CD-Bonn~\cite{Mac00}.
Taking all these differences into account,
we were able to show that
our results are consistent with the findings of Coon and Barrett.

The bottom line is that due the uncertainties in the model parameters,
there is latitude of a factor of two or so
in the strength of the $\rho$-$\omega$ potential.
Within that latitude, it is easy to fit
the full amount of CSB of the singlet scattering length.

\subsection{Nucleon mass splitting}
The difference between the masses of  neutron and proton
represents a basic cause for CSB of the nuclear force.
This source of CSB effects has been explored in great detail in ~\cite{LM98a}. 
The investigation is based upon the Bonn model for the 
NN interaction~\cite{MHE87}.
Let us briefly summarize the results.
For this we divide the total number of meson exchange diagrams that
is involved in the nuclear force into several classes. 
Below, we report the results for each class separately.

\begin{figure}
\vspace*{-3cm}
\hspace*{-1.cm}
\epsfig{file=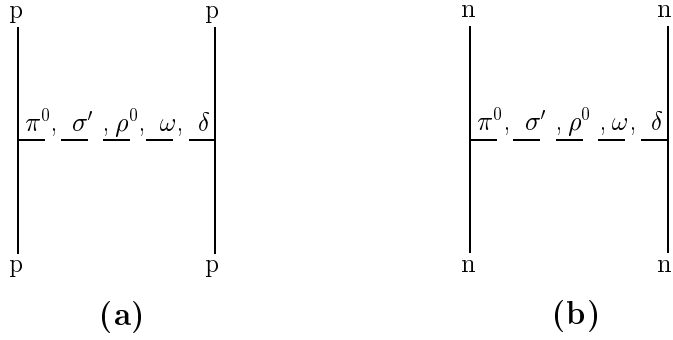,width=15cm}

\vspace*{-9cm}
\caption{One-boson-exchange (OBE) contributions to (a) $pp$
and (b) $nn$ scattering.}
\label{fig_mn1}
\end{figure}

\begin{figure}
\vspace*{-1.5cm}
\hspace*{-1.cm}
\epsfig{file=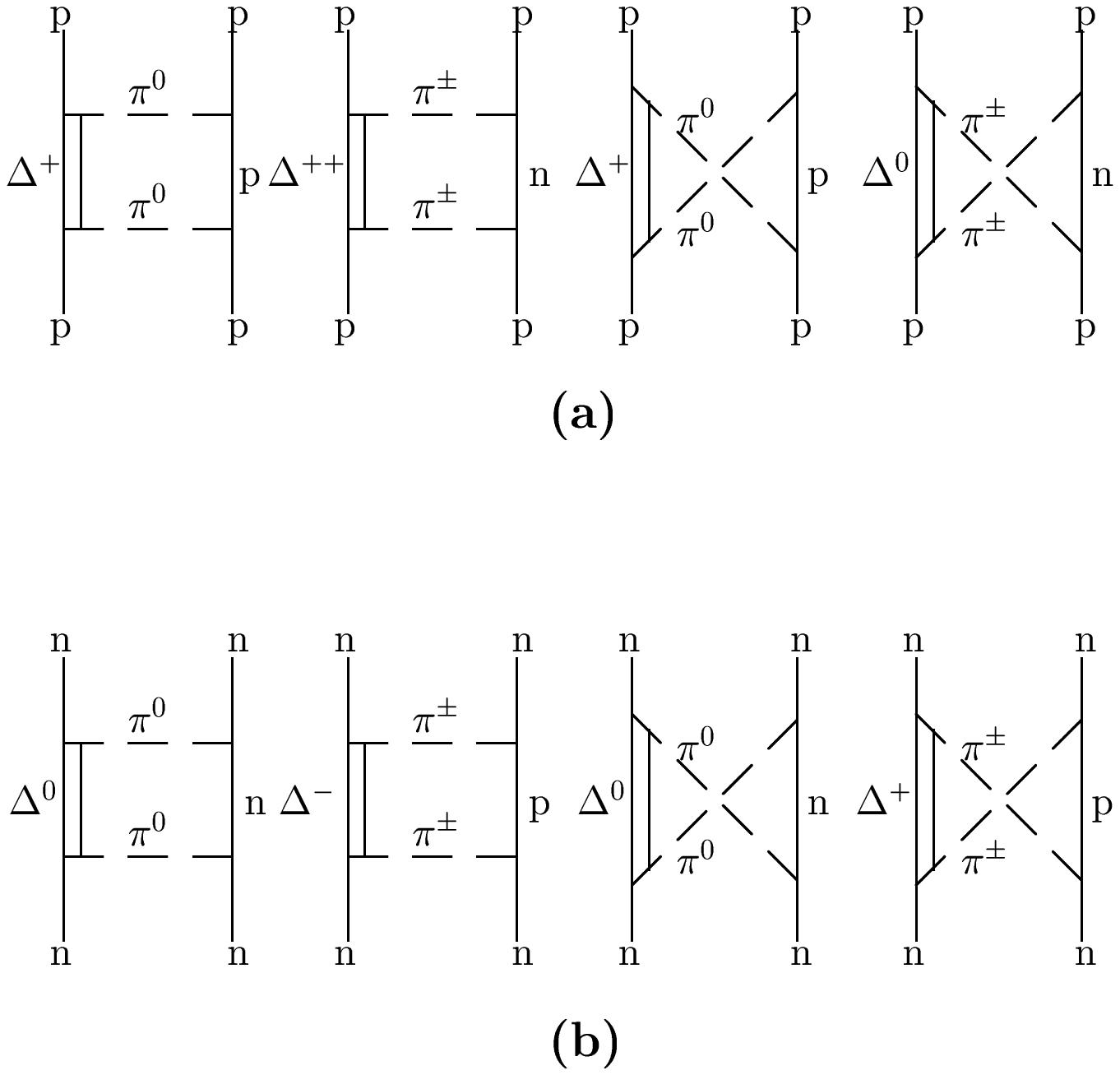,width=15cm}

\vspace*{-6.5cm}
\caption{Two-pion-exchange contributions with $N\Delta$ intermediate
states to (a) $pp$ and (b) $nn$ scattering.}
\label{fig_mn2}
\end{figure}

\begin{enumerate}
\item
{\bf One-boson-exchange} (OBE, Fig.~\ref{fig_mn1}) contributions
mediated by $\pi^{0}(135)$, $\rho^0(770)$,
 $\omega(782)$, $a_0/\delta(980)$, and
$\sigma'(550)$. 
In the Bonn model~\cite{MHE87},
the $\sigma'$ describes only the correlated $2\pi$ exchange in 
$\pi\pi-S$-wave (and not the uncorrelated $2\pi$ exchange
since the latter is calculated explicitly, cf.\ Fig.~\ref{fig_mn2}).
Charge-symmetry is broken by the fact that
for $pp$ scattering the proton mass is used in the Dirac spinors
representing the four external legs 
[Fig.~\ref{fig_mn1}(a)], 
while for $nn$ scattering the neutron mass
is applied 
[Fig.~\ref{fig_mn1}(b)]. 
The CSB effect from the OBE diagrams is very small.

\item
{\bf $2\pi$-exchange diagrams.} 
This class consists of three groups;
namely the diagrams with $NN$, $N\Delta$ and $\Delta\Delta$
intermediate states, where $\Delta$ refers to the baryon with
spin and isospin $\frac32$ and mass 1232 MeV.
The most important group is the one with $N\Delta$ intermediate
states which we show in Fig.~\ref{fig_mn2}.
Part (a) of Fig.~2 applies to $pp$ scattering, while part (b)
refers to $nn$ scattering.
When charged-pion exchange is involved, the intermediate-state
nucleon differs from that of the external legs. This is one of the
sources for CSB from this group of diagrams.
The $2\pi$ class of diagrams causes the largest CSB effect.

\item
{\bf $\pi\rho$-exchanges.} 
Graphically, the $\pi\rho$ diagrams can be obtained
by replacing in each $2\pi$ diagram (e.~g., in Fig.~\ref{fig_mn2})
one pion by a $\rho$-meson of the same charge state.
The effect is typically opposite to the one from $2\pi$ exchange.

\item
{\bf Further $3\pi$ and $4\pi$ contributions} ($\pi\sigma+\pi\omega$).
The Bonn potential also includes some $3\pi$-exchanges that can be
approximated in terms of $\pi\sigma$ diagrams and $4\pi$-exchanges
of  $\pi\omega$ type.
The sum of the two groups is small, indicating convergence of the 
diagrammatic expansion. 
The CSB effect from this class is essentially negligible.
\end{enumerate}

The total CSB difference of the singlet scattering length 
caused by nucleon mass splitting amounts
to 1.508~fm~\cite{foot} which agrees well with
the empirical value $1.6\pm 0.6$ fm.
Thus, nucleon mass splitting alone can explain the entire
empirical CSB of the singlet scattering length.

Starting from the CD-Bonn $pp$ potential~\cite{Mac00},
the parameters of the scalar-isoscalar bosons of that model
have been adjusted such that the microscopically evaluated
phase shift differences due to nucleon mass splitting are
reproduced accurately. This yields the CD-Bonn $nn$ potential.

\subsection{Phenomenological model}
An excellent example for a phenomenological construction of the CSB
nuclear force is the recent Argonne $V_{18}$ $NN$ 
potential~\cite{WSS95}.
As usual, the Argonne $pp$ potential is fixed by a best-fit to
the $pp$ data. The Argonne $nn$ potential is then constructed
by starting from the $pp$ one and readjusting the central force
in the $S=0, T=1$ state such that the empirical value for
$a_{nn}$ is reproduced. For the $S=1, T=1$ state, where 
empirical information on $nn$ scattering is not available,
it is assumed that the CSB splitting of the central force is
the same as in $S=0, T=1$.

\subsection{Comparing the predictions for the two-nucleon system}

In Table~\ref{tab_lep}, we show the $^1S_0$ effective range parameters
as calculated by the three models applied in this study.
By construction, the $\rho$-$\omega$ potential produces the
same CSB difference as the nucleon mass ($M_N$) splitting model, namely,
$\Delta a_{CSB} = 1.508$ fm. 
The Argonne $V_{18}$  potential yields
$\Delta a_{CSB} = 1.654$ fm. 
Thus, all three models have nearly identical results for
$\Delta a_{CSB}$ which is exactly what we want
as the starting point of our study.

We now turn to energies $T_{lab} > 0$ and calculate
the CSB phase shift differences
$\Delta \delta_{CSB} \equiv \delta_{nn} - \delta_{pp}$
(without electromagnetic interactions)
for all three models (see Fig.~\ref{fig_ph}).
In the $^1S_0$ state at low energies, we
have, of course, nearly identical phase shift differences
because of the agreement in $\Delta a_{CSB}$.
However, as the energy increases, differences between
the model predictions emerge. The $\rho$-$\omega$ model maintains
the largest $\Delta \delta_{CSB}$ above 150 MeV which
may be explained by the fact that $\rho$-$\omega$ exchange
is of shorter range than the $2\pi$ exchange which the $M_N$ splitting
CSB potential is based upon. The differences in $^1D_2$
can be explained with the same argument.

The largest differences between the model predictions occur
in the $^3P_J$ waves (cf.\ Fig.~\ref{fig_ph}).
As expected, the $\rho$-$\omega$ potential now
clearly reveals its large spin-orbit component
typical for vector-meson exchange. Note that the spin-orbit force of
$\Delta V_{\rho\omega}$ is of opposite sign as the one of
ordinary one-omega exchange [cf.\ Eqs.~(\ref{eq_delv}) and 
(\ref{eq_rom1}) and keep in mind that $\langle \rho^0|H_m|\omega\rangle$
is negative].

The Argonne $V_{18}$ potential follows in $^1S_0$ the
trend of the $\rho$-$\omega$ mechanism
and in $^3P_0$ it is close to the $M_N$-splitting model.
In the other partial waves, it is not close to any of the
microscopic models for CSB.

In summary, in spite of identical $\Delta a_{CSB}$,
the $\rho$-$\omega$ and the $M_N$-splitting models
show drastic differences in $^3P_J$ waves.
Unfortunately, we do not have any empirical information
on $\Delta \delta_{CSB}$ and, so, there is no direct way
to tell which is right and which is wrong.
Apart from the $^1S_0$ scattering lengths, 
the only empirical information on CSB that we have
are some binding energy differences, to
which $P$ waves do contribute.
Therefore, we will turn in the next section to such
binding energy differences with the hope that the differences
in $P$ waves may impress a detectable signature.

\section{Predictions for systems with $A>2$}

\subsection{$^3$H-$^3$He binding energy difference}

The experimental value for the difference between the binding energies
of $^3$H and $^3$He is 764 keV ($^3$H is more bound).
Most of this difference is due to
the static Coulomb energy (amongst finite-size protons)
which accounts for $648\pm 4$ keV~\cite{Bra88,FGP87,WIS90}.
Another $35\pm 3$ keV come from
electromagnetic effects neglected in the static 
Coulomb  approximation~\cite{WIS90,BCS78} 
and $14\pm 2$ keV are due to the $n-p$ mass difference in the kinetic
energy~\cite{FGP90}.
After all these obvious corrections, a binding energy difference
of $67\pm 9$ keV remains which is commonly attributed to
CSB of the nuclear force.

We have applied the three different CSB forces 
presented in the previous section
in accurate momentum-space Faddeev 
calculations of the three-nucleon systems~\cite{Bra88}.
Our results for the $^3$H-$^3$He binding energy differences
are shown in the upper part of Table~\ref{tab_app}.
We conducted two types of calculations. In the first type,
we included CSB only in $^1S_0$ while all other partial
waves are treated charge-symmetric.
In the second type, CSB was included in
all $T=1$ partial waves (i.~e., distinct $pp$ and $nn$ potentials
were used in the $T=1$ states). 

The predictions by the CSB models are between 60 and 66 keV. 
Thus, they are all consistent with the empirical value of $67\pm 9$ keV
and no discrimination is possible.
Moreover, the CSB contribution beyond $^1S_0$ is small,
2-5 keV (about 6\% of the total), which is within the empirical uncertainty.
Therefore, it is impossible to draw any conclusions concerning
the CSB contributions from $^3P_J$ waves.

The trends in the results are consistent with the
phase shift differences shown in Fig.~\ref{fig_ph}:
The $\rho$-$\omega$ model generates more binding energy
difference from the $^1S_0$ state (3.3 keV more) and more
from $^3P_J$ waves (2.5 keV more) as compared to the
$M_N$-splitting model.

\subsection{Nolen-Schiffer anomaly}

It is a well-known experimental fact that the single-particle energies
of corresponding states in mirror nuclei are different.
If one
assumes that the strong part of the nuclear force is charge symmetric, 
i.e. the strong proton-proton interaction is identical to the interaction  
between two neutrons, then these differences
would originate entirely from the electromagnetic interaction 
(mainly Coulomb) between the nucleons. 
For this reason, it is customary to call these single-particle
energy differences the Coulomb displacement energies.
After accurate experimental data on the charge distribution became
available from electron scattering experiments, Hartree-Fock calculations with
phenomenological models for the $NN$ interaction like the Skyrme forces were
performed which reproduced these measured charge distributions with good
accuracy. The Coulomb displacement energies which were evaluated with 
these Hartree-Fock wave functions, however, underestimated the experimental data
by typically seven percent. This has become known as the Nolen-Schiffer (NS)
anomaly\cite{NS69}. Many attempts have been made to explain this discrepancy 
by the inclusion of electromagnetic corrections, many-body correlations beyond
the Hartree-Fock approximation, 
or by explicit charge-symmetry breaking terms in the
NN interaction\cite{MNS90,auerb,sato,blunden,kuo}.
In these investigations it turned out that the CSB of the nuclear force
(with the $nn$ interaction more attractive than the $pp$ one)
is probably the major reason for the `anomaly'.

We will now study the impact of CSB of the $NN$ interaction on the
single-particle energies for protons and neutrons in
nuclear matter as well as in finite nuclei.

\subsubsection{Nuclear matter}
We calculate the single-particle
potential for protons, $U_p(k)$, and neutrons, $U_n(k)$,
as a function of the momentum $k$ in symmetric nuclear matter
using the self-consistent Brueckner-Hartree-Fock approach~\cite{MPM99}.
From this we obtain the CSB energy difference
\begin{equation}
\Delta U_{CSB} (k) \equiv U_p(k)-U_n(k) \, .
\end{equation}
Since the momentum dependence of $\Delta U_{CSB}$ is weak,
we choose $k=k_F$.
Our results at nuclear matter density,
$k_F=1.35$ fm$^{-1}$, are shown in the lower
part of Table~\ref{tab_app}.
The most encouraging aspect of the results is that, here,
we encounter a large contribution from states with $L>0$
(about 50\% of the total). 
Consequently, we also observe a substantial
difference between the model predictions, with the
$\rho$-$\omega$ model producing about 20\% more energy
difference than the $M_N$-splitting model.

Unfortunately, no reliable empirical estimates for $\Delta U_{CSB}$
in nuclear matter exist such that we cannot draw any conclusions.
Accurate data exist for finite nuclei which we will 
consider in the next section.

\subsubsection{Finite nuclei}

We choose $^{16}$O for our sample nucleus and we wish to
calculate Coulomb displacement energies around this 
nucleus~\cite{HMM99}.
Calculations of this kind are very involved and, therefore,
we need to discuss first how to conduct such
microscopic nuclear structure calculations in a proper way.

One possibility would be to perform self-consistent
Brueckner-Hartree-Fock (BHF) calculations and extract the Coulomb displacement
energies from the single-particle energies for protons and neutrons. We do not
take this approach, for the following reasons: (i) Such self-consistent BHF
calculations typically predict the radii for the charge-density distributions
too small\cite{carlo}. This implies
that the leading Coulomb contribution to the displacement energy would be
overestimated. Also the calculation of the correction terms would be based on
single-particle wavefunctions which are localized too much. (ii) BHF
calculations are appropriate for short-range correlations. However,
long-range correlations involving the admixture of configurations with low
excitation energies in the uncorrelated shell-model basis require a more
careful treatment. (iii) The BHF single-particle energies do not account for
any distribution of the single-particle strength consistent with
realistic spectral functions.

For the reasons listed,
we take the following approach. We use single-particle wave functions
from Hartree-Fock calculations with effective nuclear forces, which yield
a good fit to the empirical charge distribution. These wave functions
are used to determine the leading Coulomb contribution and corrections like the
effects of finite proton size, the electromagnetic spin-orbit interaction, the
kinetic energy correction due to the mass difference between proton and neutron,
and the effects of vacuum polarization. Actually, for these contributions we
use the results by Sato\cite{sato}. The first column of our Table~\ref{tab_NS}
is taken from table 2 of Ref.~\cite{sato} which includes 
all the effects just mentioned.

The correlation effects are taken into account in a two-step procedure. 
We assume a
model space defined in terms of shell-model configurations
including oscillator single-particle states up to the 1p0f shell. 
We use the oscillator parameter $b=1.76$ fm which is appropriate for
$^{16}O$. The effects of short-range correlations are calculated
by employing an effective interaction,  i.e.\ a ${\cal G}$-matrix 
suitable for the model
space. This ${\cal G}$-matrix is determined as the solution of the
Bethe-Goldstone equation
\begin{equation}
{\cal G}(\Omega ) =  V + V \frac{Q_{\hbox{mod}}}{ \Omega - Q_{\hbox{mod}} T
Q_{\hbox{mod}}} {\cal G} (\Omega )\; ,
\label{gmat}
\end{equation}
where $T$ is identified with the kinetic energy operator, while $V$
stands for the bare two-body interaction including the Coulomb interaction and
accounting for CSB in the strong interaction. The Pauli operator
$Q_{\hbox{mod}}$  in this Bethe-Goldstone Eq.~(\ref{gmat}) is defined in terms
of two-particle harmonic oscillator states $\vert \alpha\beta >$ by
\begin{equation}
 Q_{\hbox{mod}} \vert \alpha\beta > = \left\{ \begin{array}{ll}
0 & \mbox{if $\alpha$ or $\beta$ from $0s$ or $0p$ shell}\cr
0 & \mbox{if $\alpha$ and $\beta$ from $1s0d$ or $1p0f$ shell} \cr
\vert \alpha\beta > & \mbox{elsewhere} \end{array} \right.
\label{paul}
\end{equation} 
As a first approximation we use the resulting ${\cal G}$-matrix elements and
evaluate single-particle energies in the BHF approximation
$\epsilon_{\alpha}$. This
approximation, which will be denoted as BHF in the discussion below, 
accounts for short-range correlations, which are described in terms of
configurations outside our model space. In a next step we add to this BHF
definition of the nucleon self-energy the irreducible terms of second order in
${\cal G}$ which account for intermediate two-particle one-hole and one-particle
two-hole configurations within the model-space
\begin{equation}
{\cal U}_{\alpha}^{(2)} = \frac{1}{2} \sum_{p_1,p_2,h}\frac{<\alpha h \vert
 {\cal G}\vert p_1p_2><p_1p_2\vert {\cal G}\vert \alpha h>}{\omega -
(\epsilon_{p_1} + \epsilon_{p_2}-\epsilon_h) + i\eta} +
\frac{1}{2} \sum_{h_1,h_2,p}\frac{<\alpha p \vert
 {\cal G}\vert h_1h_2><h_1h_2\vert {\cal G}\vert \alpha p>}{\omega -
(\epsilon_{h_1} + \epsilon_{h_2}-\epsilon_p) - i\eta}\,.
\end{equation}
 Applying the techniques
described in \cite{skour}, we can solve the Dyson equation for the
single-particle Greens function $G_{\alpha}(\omega )$
\begin{equation}
G_{\alpha}(\omega) = g^{\alpha}(\omega) + 
g_{\alpha}(\omega){\cal U}_{\alpha}^{(2)}(\omega) G_{\alpha}(\omega)
\end{equation}
with $g_\alpha$ the BHF approximation for the single-particle Greens function, 
and determine its Lehmann representation
\begin{equation}  
G_{\alpha}(\omega ) = \sum_n \frac{\left| <\Psi_n^{A+1} \vert 
a^\dagger_\alpha \vert \Psi_0 >\right|^2} {\omega - (E_n^{A+1}-E_0) + i\eta}
+  \sum_m \frac{\left| <\Psi_m^{A-1} \vert 
a_\alpha \vert \Psi_0 >\right|^2} {\omega - (E_0 -E_m^{A-1}) - i\eta}\,.
\label{lehm}
\end{equation}
This yields directly the energies of the states with $A\pm 1$ nucleons 
that we are interested in.

Our results for the Coulomb displacement energies 
are listed in Table~\ref{tab_NS} for various one-hole and one-particle states
relative to $^{16}$O. The first column of this table, $C^{(1)}$,
contains the results of
Ref.~\cite{sato} for the leading Coulomb contributions, 
the corrections due to the
finite proton size, the electromagnetic spin-orbit interaction, the
kinetic energy correction due to nucleon mass splitting,
and the effects of vacuum polarization. As discussed above, we think
that it is more realistic to evaluate these contributions for single-particle
wave functions which are derived from Hartree-Fock calculations with
phenomenological forces rather than using the wavefunctions derived from a
microscopic BHF calculation.

The second and third columns of Table~\ref{tab_NS} 
list the corrections to the Coulomb
displacement energies which originate from the treatment of short-range
($\Delta_{SR}$) and long-range correlations ($\Delta_{LR}$) discussed above. The
correction $\Delta_{SR}$ has been derived from the differences of
BHF single-particle energies for protons and neutrons subtracting the Coulomb
displacement energy evaluated in the mean-field approximation
\begin{equation}
\Delta_{SR} = \epsilon_i^{BHF} (\mbox{proton}) - \epsilon_i^{BHF}
(\mbox{neutron}) - \Delta_{\mbox{mean field}} 
\end{equation}
In this case the BHF calculations have been performed with the $np$ versions of
the different interactions. The correction terms
$\Delta_{LR}$ have been evaluated in a similar way from the quasiparticle
energies determined in the Greens function approach, subtracting the BHF effects
already contained in $\Delta_{SR}$. The correction terms $\Delta_{SR}$ and
$\Delta_{LR}$ include the effects represented by irreducible diagrams of second
and higher order in the interaction, in which at least one of the interaction
lines represents the Coulomb interaction. In addition they contain the effects 
of folded diagrams discussed by Tam et al.\cite{kuo}. We find that the
correlation effects are rather weak. The short- and long-range contributions
tend to cancel each other. This is true in particular for the one-hole states
$p_{3/2}^{-1}$ and $p_{1/2}^{-1}$. The effects of short-range correlations
dominate in the case of the particle states, $d_{5/2}$ and $1s_{1/2}$, leading
to a total correlation effect in the order of 100 keV 
in the Coulomb displacement
energies. This effect is slightly larger for the Argonne potential than for the
Bonn potentials because of the stronger correlations in the case of Argonne.

The contributions to the Coulomb displacement energies caused by CSB
of the $NN$ interactions, $\Delta_{CSB}$, 
are listed in the fourth column of Table~\ref{tab_NS}.
We have conducted separate calculations for each of the
three models for CSB introduced in Sec.~II.

We note that, also in the calculations of the Coulomb displacement
energies, it is important to include CSB beyond the 
$^1S_0$ state in the $NN$ interaction.
Similarly to what we found for the nuclear matter $\Delta U_{CSB}$
(cf.\ Sec.~III.B.1 and lower part of Table~\ref{tab_app}),
CSB in $P$ and higher partial waves contributes about 50\%
of the total $\Delta_{CSB}$. This was demonstrated in Ref.~\cite{HMM99}.

We achieve satisfactory or even good agreement in some cases, 
like $p^{-1}_{3/2}$ and $1s_{1/2}$, 
but there are discrepancies in others.
For $p^{-1}_{1/2}$ and $d_{5/2}$ the remaining discrepancies
are larger than the $\Delta_{CSB}$ contribution.

The general trend in the results is that the $\rho$-$\omega$ model
generates about 10-20\% more $\Delta_{CSB}$ than the $M_N$-splitting
mechanism, and Argonne $V_{18}$ is, in general, in-between the two.
Even though the $\rho$-$\omega$ trend is a favourable one, 
it is not sufficiently pronounced such that one could
give a preference to this model.
In the critical cases, like 
$p^{-1}_{1/2}$ and $d_{5/2}$,
the discrepancies between all model predictions, on the one hand,
and experiment, on the other, 
are much larger
than the differences within the model predictions.

Concerning the remaining discrepancy, a comment is in place.
Note that the nuclear structure part of our calculations may carry
some uncertainty. This is true in particular for the evaluation of the leading
Coulomb contribution, which is sensitive to the Hartree-Fock wave functions
which are used. To obtain an idea of how large such uncertainties
may be, we compare the results for Coulomb displacement energies
using the Skyrme II force and {\it no CSB} by Sato~\cite{sato}
with the more recent ones by Suzuki {\it et al.}~\cite{SSA92}.
For the single-hole state $p^{-1}_{1/2}$, 
Suzuki's result is larger by 167 keV as compared to Sato; 
and for the single-particle state $d_{5/2}$, the two calculations
differ by 138 keV.
Uncertainties of this size can well explain the remaining discrepancies
in our results.

\section{Summary and conclusions}

We have tested three different models for the charge symmetry breaking (CSB)
of the $NN$ interaction.
The models are based upon $\rho$-$\omega$ mixing, nucleon mass ($M_N$)
splitting, and phenomenology (Argonne $V_{18}$).
All models reproduce the empirical value for the CSB of
the $^1S_0$ scattering length ($\Delta a_{CSB}$)
accurately. 

We reveal that there are considerable differences in the
predictions by these models for CSB in $^3P_J$ waves.
We have investigated the impact of these differences on some
observable quantities of $A\geq 3$ nuclear systems
that are sensitive to CSB of the nuclear force.

We find that the $^3$H-$^3$He binding energy difference
is essentially ruled by $\Delta a_{CSB}$ and that $P$ and
higher partial waves contribute only about 6\%.
Therefore, this quantity is unsuitable to discriminate between
different models for CSB of the nuclear force.

A test calculation conducted for nuclear matter
shows that in heavier nuclear systems
the difference between proton and neutron single-particle
energies receives about 50\% from partial waves other
than $^1S_0$. Motivated by this result, we have calculated
the Coulomb displacement energies around the closed shell
nucleus $^{16}$O. We find that the contribution
to these energy differences from CSB of the $NN$ interaction
(which is in the order of 100 keV) differs by 10-20\%
among the three models for CSB, which is appreciable.
Unfortunately, the nuclear structure part of these calculations, in particular
the evaluation of the leading Coulomb contributions,
carry an uncertainty in the order of 100 keV such that
the subtle differences between the competing CSB models 
get swamped. Therefore we must conclude that, based upon the calculations
conducted in this study, we are unable to give
preference to any of the three CSB models.

What we need are observables for which the nuclear structure
part is fully under control. This suggests to look into nuclear
few-body reactions for which exact calculations can
be performed~\cite{Glo96}. An example could be the analyzing power, $A_y$,
in nucleon-deuteron scattering that is known to depend
sensitively on the $^3P_J$ waves of the $NN$ potential.
Accurate data on $p-d$ and $n-d$ $A_y$ exist and these
data exhibit a clear signature of CSB. Unfortunately, these
data cannot be explained without resource to three-body forces
which may obscure the CSB aspect of the problem.
In any case, we like to encourage the nuclear few-body
community~\cite{Glo96} to identify spin observables in few-body reactions
that show sensitivity to CSB of the $NN$
interaction, particularly, the one that comes
from $^3P_J$ waves.
Investigations of this kind may ultimately allow to
discriminate between competing models for CSB of
the nuclear force.

\acknowledgments
This work was supported in part by the U.S. National Science Foundation
under Grant No.~PHY-9603097 and by the German DFG (SFB 382).

\begin{table}
\caption{$^1S_0$ scattering length ($a$) and effective range ($r$), for
proton-proton ($pp$) and neutron-neutron ($nn$), with Coulomb effects
($C$) and without any electromagnetic effects ($N$),  
in units of fm.}
\begin{tabular}{ldddc}
                   
 & $\rho$-$\omega$ mixing
 & $M_N$ splitting
 & Argonne $V_{18}$
 & Experiment
\\
\hline 
$a_{pp}^C$ & --7.8154& --7.8154& --7.8138  & $ -7.8149 \pm 0.0029^{a} $ \\
$r_{pp}^C$ &  2.773   & 2.773 & 2.787 & $2.769 \pm 0.014^a$ \\
$a_{pp}^N$ &--17.460  & --17.460 & --17.164  &                 \\
$r_{pp}^N$ &  2.845   & 2.845 & 2.865 &                        \\
$a_{nn}^N$ &--18.968  & --18.968 & --18.818 &$-18.9 \pm 0.4^b$  \\
$r_{nn}^N$ &  2.816   & 2.819 & 2.834 & $2.75 \pm 0.11^c$ \\
$\Delta a_{CSB}$ & 1.508 & 1.508 & 1.654 & $1.6 \pm 0.6$ \\
$\Delta r_{CSB}$ & 0.029 & 0.026 & 0.031 & $0.10 \pm 0.12$\\
\end{tabular}
$^a$Reference~\cite{SES83}.\\
$^b$Reference~\cite{How98,Gon99}.\\
$^c$Reference~\cite{MNS90}.
\label{tab_lep}
\end{table}

\vspace*{3cm}

\begin{table}
\caption{Applications of CSB potentials in the three-nucleon system
and in symmetric nuclear matter.}
\begin{tabular}{ldddc}
                   
 & $\rho$-$\omega$ mixing
 & $M_N$ splitting
 & Argonne $V_{18}$
 & Empirical
\\
\hline 
\hline 
\multicolumn{5}{c}{\bf $^3$H-$^3$He binding energy difference (keV)}\\
CSB in $^1S_0$ only & 60.9 & 57.6 & 62.1 &         \\
CSB in all $T=1$ states & 65.8 & 60.0 & 65.1 & 67 $\pm$ 9 \\
\hline
\multicolumn{5}{c}{\bf Nuclear matter $\Delta U_{CSB}$ (MeV)}\\
CSB in $^1S_0$ only & 0.168 & 0.154 & 0.180 &      \\
CSB in all $T=1$ states & 0.367 & 0.311 & 0.301 &   \\
\end{tabular}
\label{tab_app}
\end{table}

\begin{table}
\caption{ Coulomb displacement energies for single-hole ($p_{3/2}^{-1}$ and 
$p_{1/2}^{-1}$) and single-particle states ($d_{5/2}$ and $1s_{1/2}$) around
$^{16}$O. The single-particle contribution, $C^{(1)}$, is from Sato
\protect\cite{sato}. Contributions due to short-range correlations,
$\Delta_{SR}$, long-range correlations inside the model space, $\Delta_{LR}$,
and due to CSB of the strong interaction,
$\Delta_{CSB}$, are
calculated for three different CSB models. 
The total results for the displacement energies, $C^{Tot}$, are compared to
the experimental data given in the last column. All entries are in keV.}
\begin{tabular}{cc|rrrrrr}
&& $C^{(1)}$ & $\Delta_{SR}$ & $\Delta_{LR}$ & $\Delta_{CSB}$ & $C^{Tot}$ & 
Exp \\
\hline
$p_{3/2}^{-1}$ & $\rho$-$\omega$ mixing & 3205 &  -44 & 46 & 106 & 3313 & 3395 \\
& $M_N$ splitting & & -44 & 46 & 97 & 3303 & \\
& Argonne $V_{18}$ & & -71 & 47 & 108 & 3285 & \\
\hline
$p_{1/2}^{-1}$ & $\rho$-$\omega$ mixing & 3235 & -52 & 37 & 124 & 3344 & 3542 \\
& $M_N$ splitting & & -52 & 37 & 102 & 3322 & \\
& Argonne $V_{18}$ & & -79 & 39 & 103 & 3297 & \\
\hline
$d_{5/2}$ & $\rho$-$\omega$ mixing & 3135 & 154 & -15 & 93 & 3367 & 3542 \\
& $M_N$ splitting & & 154 & -15 & 87 & 3361 & \\
& Argonne $V_{18}$ & & 187 & -18 & 92 & 3401 & \\
\hline 
$1s_{1/2}$ & $\rho$-$\omega$ mixing & 2905 & 159 & -45 & 134 & 3154 & 3166 \\  
& $M_N$ splitting & & 160 & -46 & 112 & 3132 & \\
& Argonne $V_{18}$ & & 198 & -47 & 112 & 3174 & \\
\end{tabular}
\label{tab_NS}
\end{table}

\vspace*{3cm}

\begin{figure}
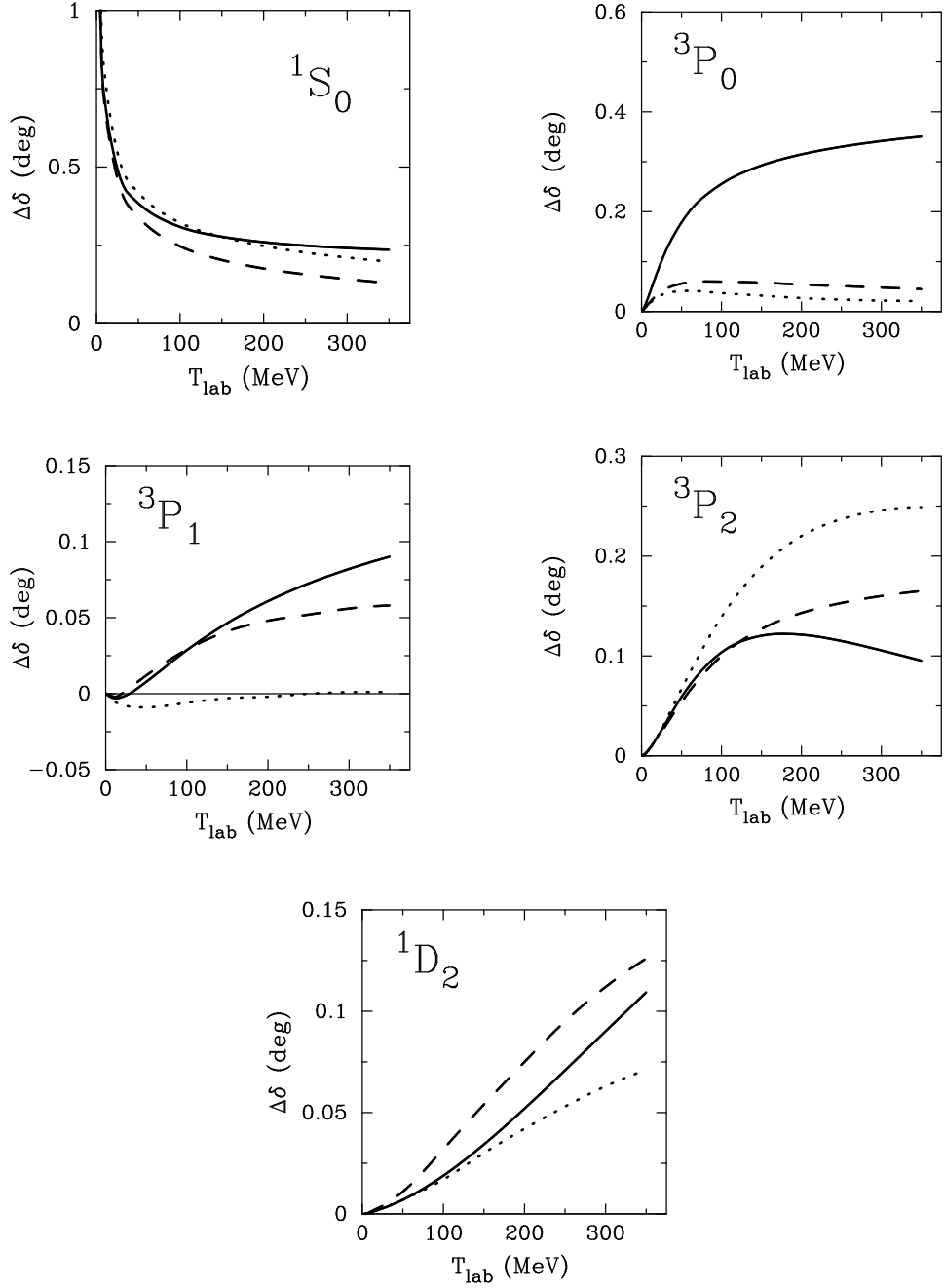


\vspace{2.0cm}
\hspace*{0.50cm}
\epsfig{file=romfig1.ps,width=5.5cm}

\vspace*{-5.4cm}
\hspace*{7.75cm}
\epsfig{file=romfig2.ps,width=5.5cm}

\vspace*{1.00cm}
\hspace*{0.50cm}
\epsfig{file=romfig3.ps,width=5.5cm}

\vspace*{-5.4cm}
\hspace*{7.75cm}
\epsfig{file=romfig5.ps,width=5.5cm}

\vspace*{1.00cm}
\hspace*{4.00cm}
\epsfig{file=romfig4.ps,width=5.5cm}

\vspace*{1.0cm}
\caption{CSB phase shift differences $\delta_{nn}-\delta_{pp}$ 
(without electromagnetic interactions)
for laboratory kinetic energies $T_{lab}$
below 350 MeV and partial waves with
$L\leq 2$ as generated by
$\rho$-$\omega$ mixing (solid line),
nucleon mass splitting (dashed),
and the phenomenological Argonne $V_{18}$ model (dotted).}
\label{fig_ph}
\end{figure}


\begin{references}
\bibitem{foot} 
Based upon the Bonn Full Model~\cite{MHE87},
the original investigation~\cite{LM98a} of the effect of nucleon mass 
splitting
uses the $\pi NN$ coupling constant $g^2_\pi/4\pi=14.4$ 
which is not current. For that reason we
have revised the Bonn Full Model using
$g^2_\pi/4\pi=13.6$ and then repeated the CSB calculations
of Ref.~\cite{LM98a}. The total $\Delta a_{CSB}$ predicted
by the revised model is 1.508 fm (about 5\% less than what was
reported in Ref.~\cite{LM98a}). Similarly, the phase shift differences
created by nucleon mass splitting shown in Fig.~\ref{fig_ph}
by the dashed curve are, in general, a few percent smaller than those
given in Ref.~\cite{LM98a}.
\end{references}
\end{document}